\documentclass[10pt]{article}
\usepackage{blois,epsfig}

\def\Journal#1#2#3#4{{#1} {\bf #2}, #3 (#4)}

\def\NPB{{\em Nucl. Phys.} B}
\def\PLB{{\em Phys. Lett.}  B}

\begin{document}
\begin{flushleft}
{\it 23rd Recontres de Blois\\ 
29 May -- 3 June 2011\\ 
Blois, France} \hfill TTK-11-49
\end{flushleft}
\vspace*{2cm}
\title{NNLO analysis of the LHC $W$ lepton charge asymmetry data}

\author{ M. Ubiali }

\address{Institut f\"ur Theoretische Teilchenphysik und Kosmologie,\\ RWTH 
Aachen University, D-52056 Aachen, Germany}

\maketitle\abstracts{
The reweighting method presented in earlier publications is applied for incorporating the LHC W lepton asymmetry data published in 2010 into the NNPDF2.1 NNLO analysis. We confirm the result of the NLO analysis which indicated that these data reduce PDF uncertainties of light quarks in the medium and small--$x$ region, providing the first solid constraints on PDFs from LHC data.}

\section{Introduction}

The knowledge of PDFs and of their associate uncertainties plays a crucial role in the LHC phenomenology, particularly in a situation where the error on PDFs represents the dominant source of uncertainty for several key processes. On the other hand the LHC itself is providing the PDF fitting collaboration with a large amount of precise data thanks to the high statistics accumulated and the well--controlled systematics. In a short--medium term, LHC measurements are going to provide essential constraints on most PDF combinations.

In a series of previous paper~\cite{nnpdf10,nnpdf12,nnpdf20,nnpdf21,nnpdf21nnlo} we have presented a new method for extracting PDFs from experimental data based on the combination of a Monte Carlo sampling technique and the use of neural networks as unbiased parametrization. The NNPDF parton sets provide a Monte Carlo representation of the probability density in the space of PDFs. Such feature enables us to include new information provided by new experimental data by using Bayes'theorem, i.e. by reweighting an existing NNPDF set (prior probability) without having to perform a new fit. The validity of such technique was demonstrated in a set of previous publications~\cite{rw,rwLHC}. The reweighting method was first applied to study the compatibility and the impact of the D0 $W$ lepton charge asymmetry data on PDFs~\cite{rw}. It was then employed~\cite{rwLHC} to assess the effect of the inclusion of the $W$ lepton charge asymmetry measurements collected at the LHC in 2010 on a global set of PDFs. The prior probability was provided by the NNPDF2.1 set~\cite{nnpdf21}. The latter is a NLO determination of parton distributions from a global set of hard scattering data using the NNPDF methodology that includes the heavy quark mass effects through the FONLL General Mass Variable Flavor Number scheme~\cite{FONLL}.

Recently the LO and NNLO fits based on the same set of experimental data and theory were published~\cite{nnpdf21nnlo}. The NNLO set is needed for the evaluation of LHC standard candle processes, which in some cases are characterized by large NNLO QCD corrections. Even though QCD radiative corrections to the $W$ lepton asymmetry were shown to be small~\cite{grazzini} (but not as small as those to the W charge asymmetry), and therefore we do not expect to observe any significant change in the results of the reweighting analysis, it is anyway interesting to re--perform it at NNLO. It confirms its perturbative stability and provides the companion NNLO parton set of the NLO NNPDF2.2 set that was recently presented~\cite{rwLHC}.

\section{Inclusion of $W$ lepton charge asymmetry data in the NNPDF2.1 NNLO analysis}

The LHC measuremets that we include in the NNLO NNPDF2.1 fit are the lepton charge asymmetry data collected and published by ATLAS~\cite{atlas} and CMS~\cite{cms} collaborations in 2010~\footnote{In future analyses the more recent data published by ATLAS~\cite{Aad:2011dm} and the ones which are to be released soon by the CMS collaboration will substitute the data considered in this work. The new data, being more precise and including a full covariance matrix, supersede previous measurements.}. The ATLAS muon asymmetry measurement is based on 31pb$^{-1}$ of accumulated luminosity in the pseudorapidity range $|\eta|<2.4$. CMS presented data for both the electron and muon charge asymmetries from W decays with two different cuts on the transverse momentum of the detected lepton: $p_{\bot}>$ 25 GeV and $p_{\bot}>$ 30 GeV. They are based on 36pb$^{-1}$ of accumulated luminosity, in the pseudorapidity range $|\eta|<2.2$. Here we only consider the dataset with the looser cut $p_{\bot}>$ 25 GeV. These datasets were shown to provide a constraint for light quark and antiquark in the $10^{-3} < x < 5\cdot 10^{-2}$, where they are only partially constrained by the data already included in the NNPDF global analysis. In particular, while $u$ is well determined by fixed target DIS data, $d$ and the light sea combination $(\bar{d}-\bar{u})$ are much less constrained.

\begin{figure}
\epsfig{figure=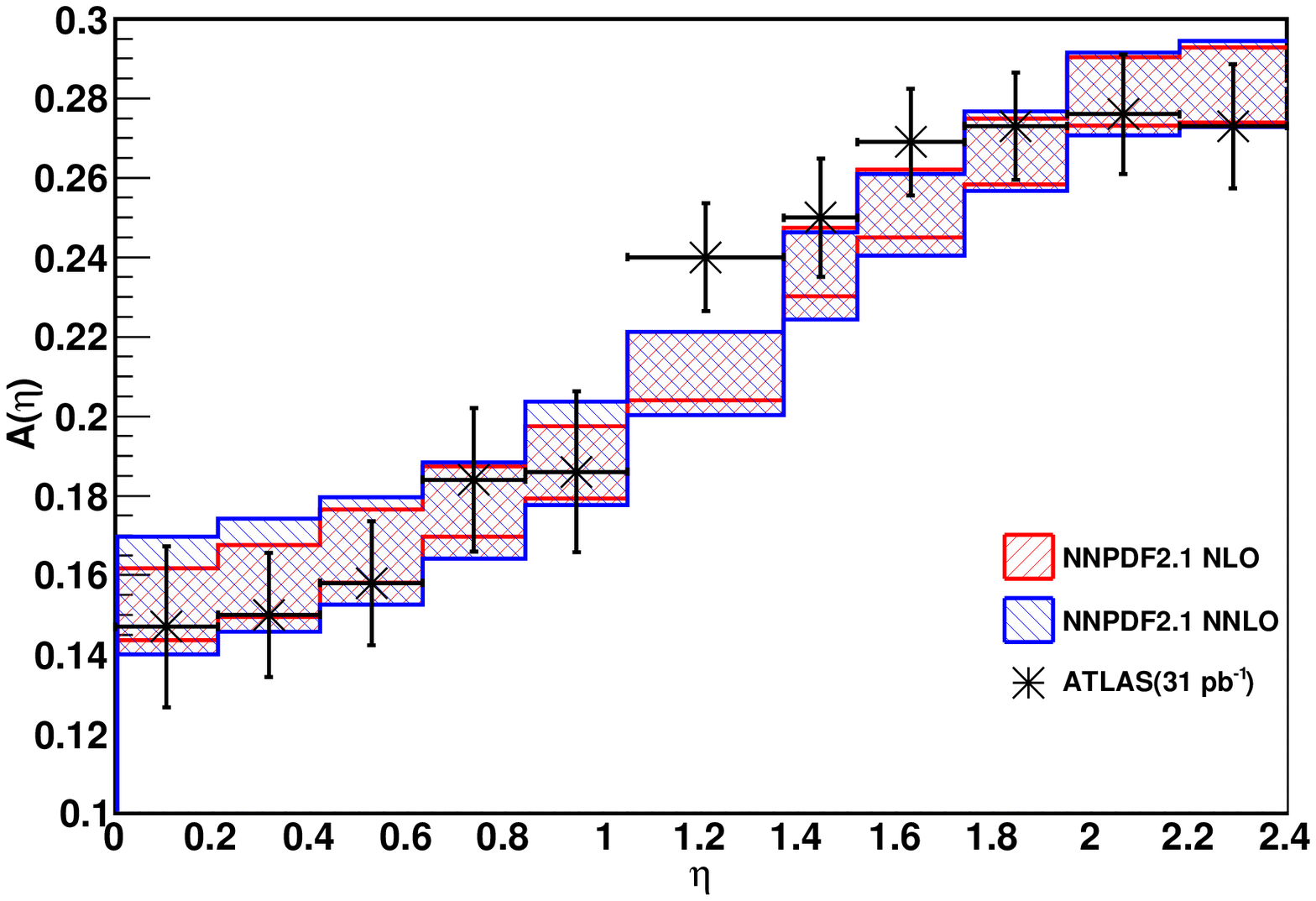,width=0.32\textwidth}
\epsfig{figure=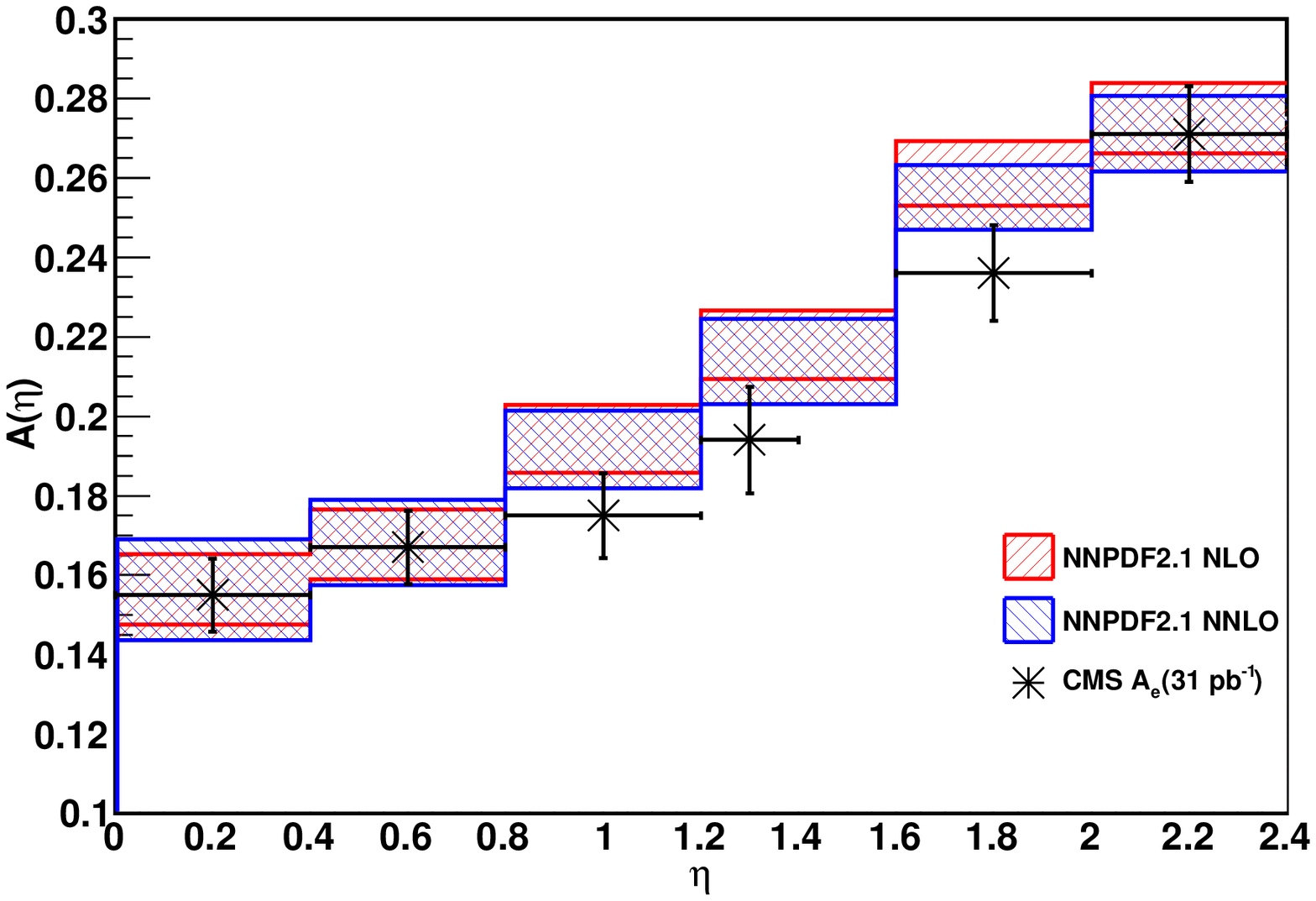,width=0.32\textwidth}
\epsfig{figure=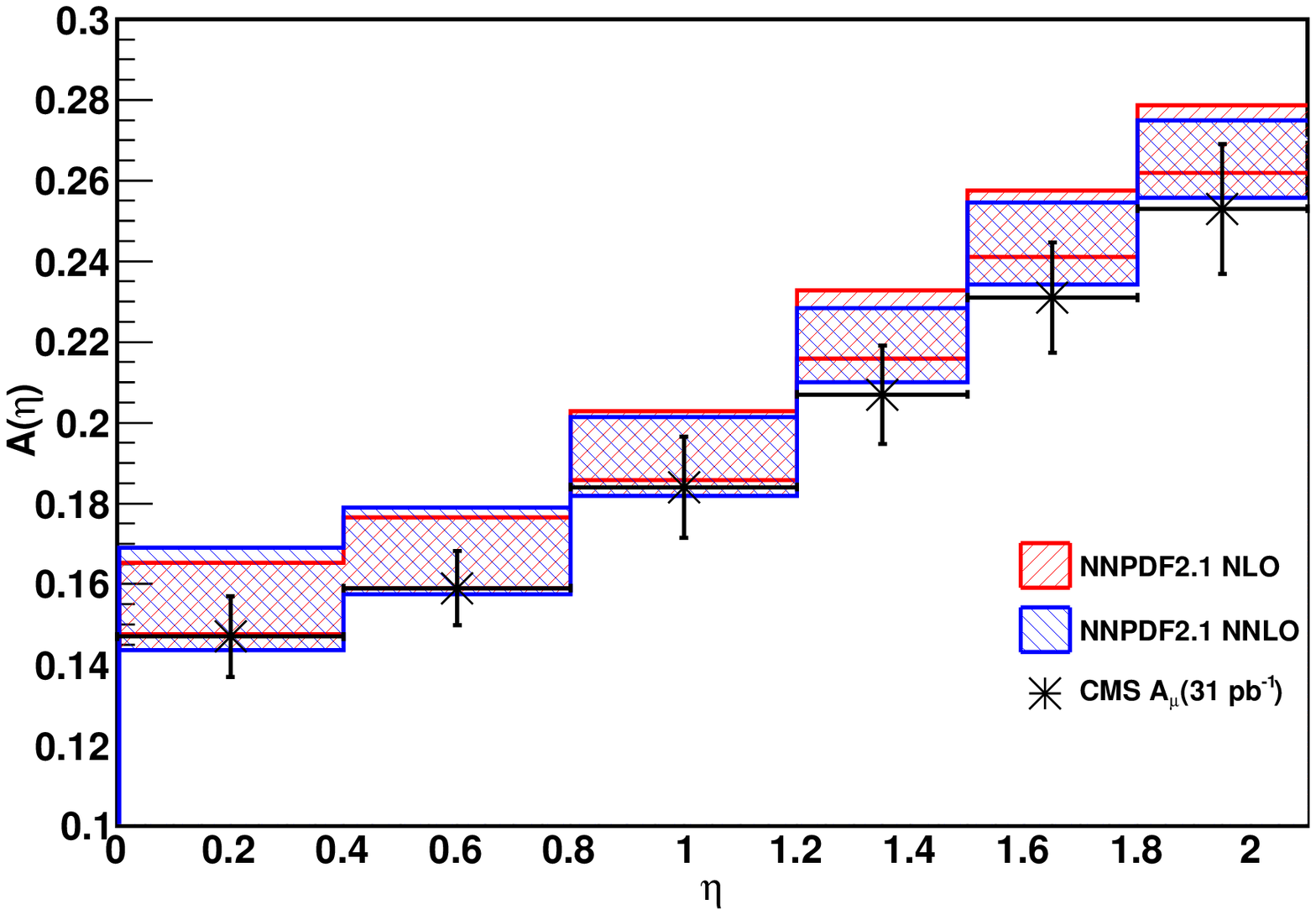,width=0.32\textwidth}
\caption{\label{fig:pred}Predictions for the $W$ lepton asymmetry at NLO and NNLO, obtained with DYNNLO~\protect\cite{dynnlo} using the NNPDF2.1 NLO and NNPDF2.1 NNLO sets respectively, compared to measurements for the muon charge asymmetry from ATLAS~\protect\cite{atlas} (left), and the electron (centre) and muon (right) charge asymmetries from CMS~\protect\cite{cms}.}
\end{figure}

In Fig.~\ref{fig:pred} we compare the NLO and NNLO predictions obtained using the fully differential Monte Carlo code DYNNLO~\cite{dynnlo}
which allows for the implementation of arbitrary experimental cuts. A more quantitative estimate of their level of agreement  with the experimental data, is shown in Table~\ref{tab:chi2} where the $\chi^{2}/{\rm d.o.f.}$ is provided for each individual dataset before and after its inclusion in the fit. 
The ATLAS muon charge asymmetry data are already very well described by the NNPDF2.1 prediction, both at NLO and at NNLO, before being included in the analysis. The description of the CMS data instead improves noticeably when comparing the NLO and the NNLO predictions. The uncertainty of the prediction is basically unchanged but the central values move closer to the experimental measurements, especially in the higher rapidity bins.
\begin{table}[t]
\caption{Values of $\chi^{2}/{\rm d.o.f.}$ for lepton charge asymmetry data for the NLO and NNLO NNPDF parton sets before (in italic) and after the inclusion in the fit. Theory predictions are computed at NLO and NNLO accuracy using the DYNNLO code. Since no covariance matrix is provided for these data, statistical and systematic uncertainties are added in quadrature in the computation of the $\chi^{2}$.\label{tab:chi2}}
\vspace{0.4cm}
{\small
\begin{center}
\begin{tabular}{|c|c||c|c||c|c|}
\hline
& $N_{\rm dat}$ & NNPDF2.1 nlo & NNPDF2.1 nnlo & NNPDF2.2 nlo & NNPDF2.2 nnlo\\
\hline
ATLAS  & 11 & {\it 0.76} & {\it 0.85}  & 1.07  & 0.90 \\
CMS e &  6 & {\it 1.83} & {\it 1.19} & 1.08  & 0.66 \\
CMS $\mu$ &  6 & {\it 1.24} & {\it 0.80} & 0.56  & 0.34 \\
\hline
D0 e  & 12 & {\it 4.39} & {\it 2.79} & 1.38 & 1.85 \\
D0 $\mu$  & 10 & {\it 1.48} & {\it 1.93} & 0.35 & 0.54 \\
\hline
\end{tabular}
\end{center}}
\end{table}

\begin{figure}
\epsfig{figure=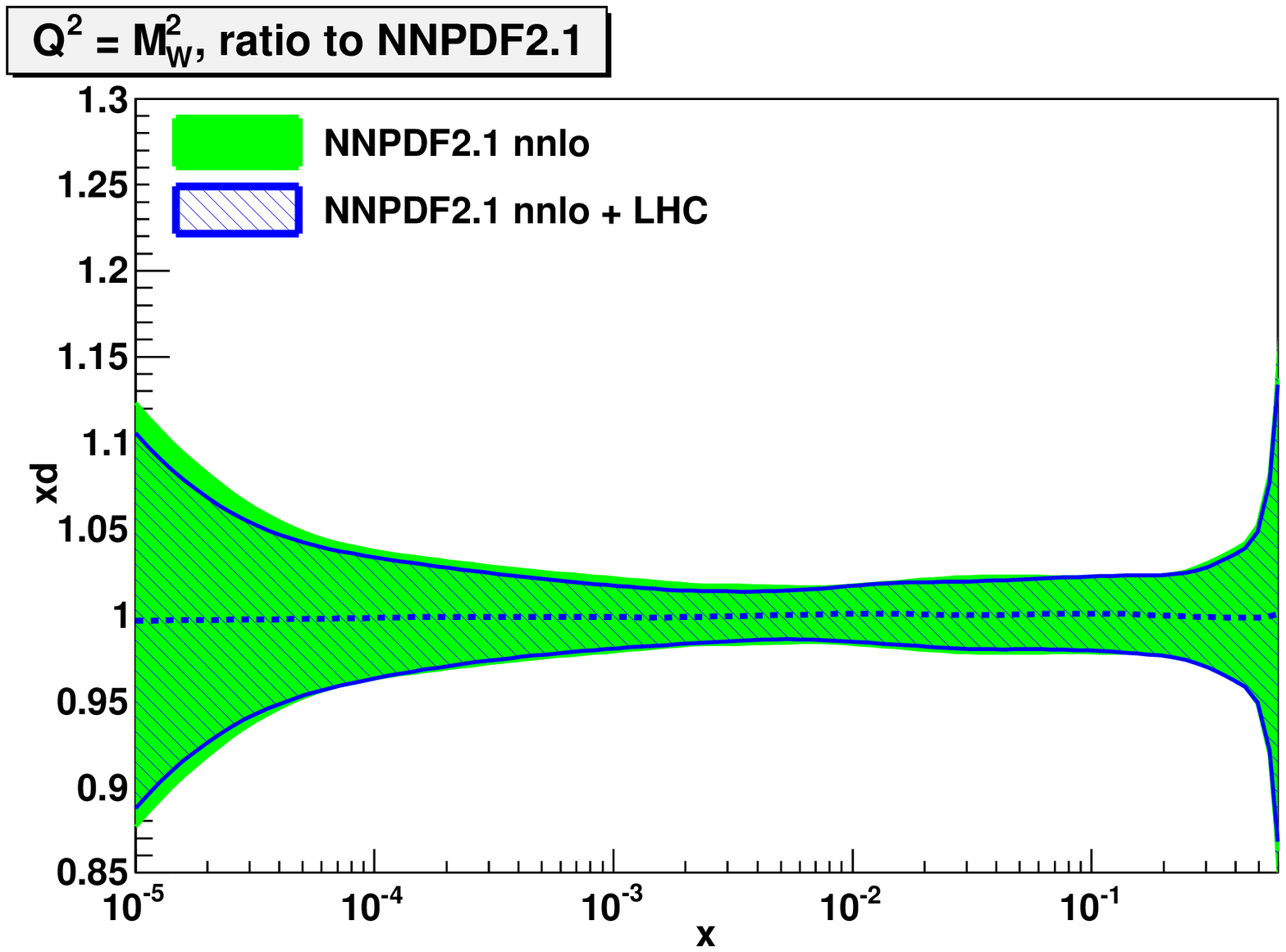,width=0.48\textwidth}
\epsfig{figure=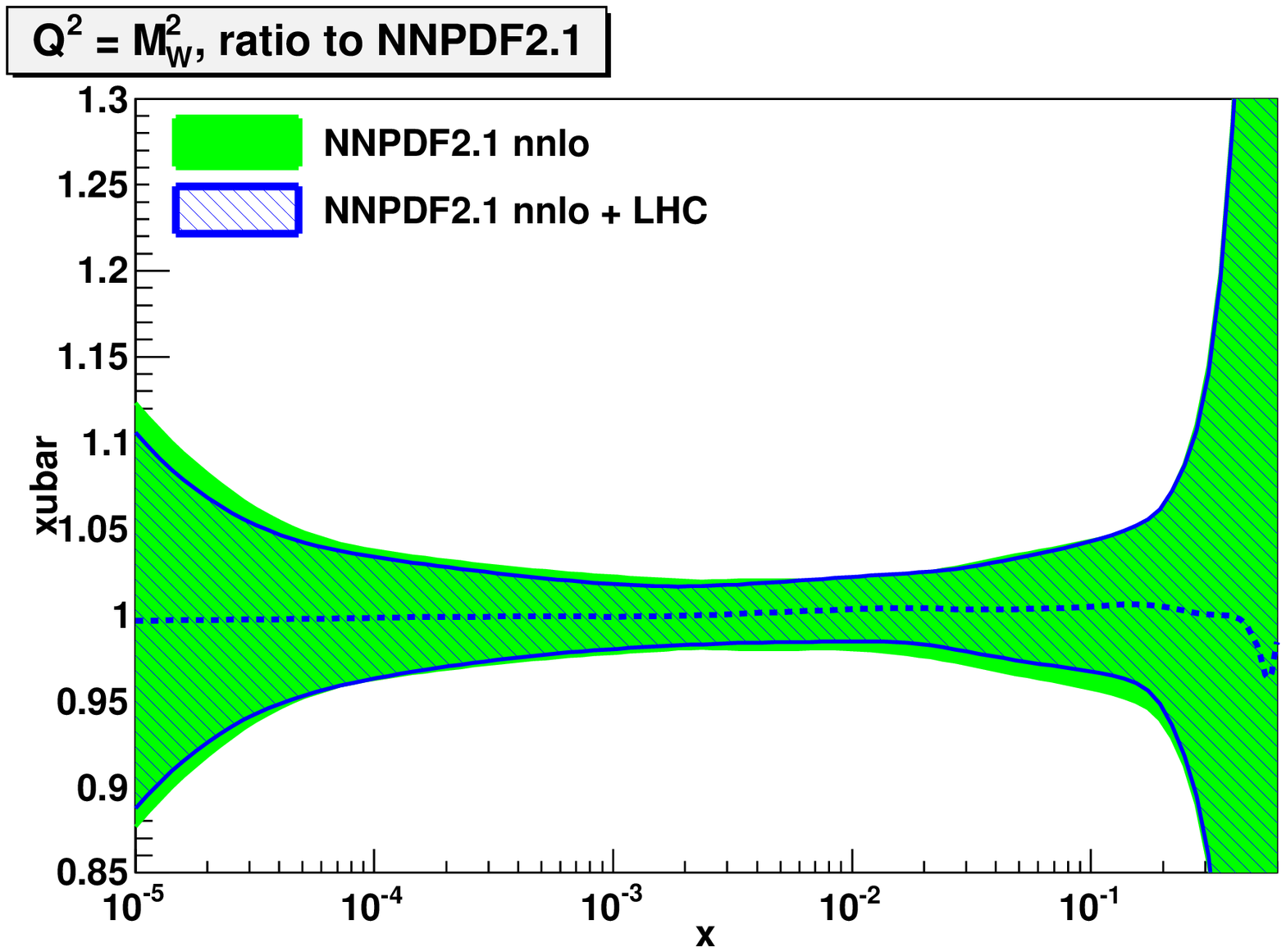,width=0.48\textwidth}
\caption{ Comparison of light quark and antiquark distributions at the scale $Q^{2}=M_{W}^{2}$ from the global NNPDF2.1 NNLO global fit and the same distributions obtained after adding ATLAS and CMS charge asymmetry data via reweighting. Parton densities are plotted normalized to the NNPDF2.1 NNLO central value\label{fig:pdfLHC}}
\end{figure}

We first add the ATLAS and CMS lepton charge asymmetry data as a single dataset to the NNPDF2.1 NNLO global fit using reweighting.
The whole dataset is already well described by the NNPDF2.1 NNLO parton set with $\chi^{2}/{\rm d.o.f.}$ = 0.93 and a distribution  for individual replicas having a sharp peak around one. After reweighting the description of the data improves, with $\chi^{2}/{\rm d.o.f.}$ = 0.78. These results, combined with the number of effective replicas surviving after reweighting, namely $N_{\rm eff}$ = 669 out of the initial $N_{\rm rep}$ = 1000, show that the use of the ATLAS and CMS data together in the fit does not cause any issue and imposes a moderate constraint on light quark PDFs, as it is shown in Fig.~\ref{fig:pdfLHC}.
There is around 20\% reduction in uncertainties at small--medium $x$ values, analogous to the reduction observed at NLO~\cite{rwLHC}.

In a previous study~\cite{rw} we have shown that the Tevatron D0 lepton charge asymmetry data that are inclusive in the $p_{\bot}$ the identified lepton, namely the muon charge asymmetry~\cite{d0mu} and electron charge asymmetry data with $p_{\bot}>$ 25 GeV~\cite{d0el}, are consistent with each others and with all the other datasets included in NNPDF analysis, in particular with the CDF W asymmetry data~\cite{cdf} and the fixed-target DIS deuteron data. Less inclusive electron charge asymmetry data, binned in $p_{\bot}$, were shown to be inconsistent with some of the DIS data included in the global analysis and have problems of internal consistency. We have then excluded these datasets. In Refs.~\cite{rw,rwLHC} the muon charge asymmetry and inclusive electron charge asymmetry data were shown to provide additional information to that coming from the LHC measurements. Therefore, after checking the consistency of the Tevatron D0 data with the LHC one and with the data already included in the NNLO global analysis, we proceed directly to a combined fit of these data together with the LHC data.

\begin{figure}[h]
\epsfig{figure=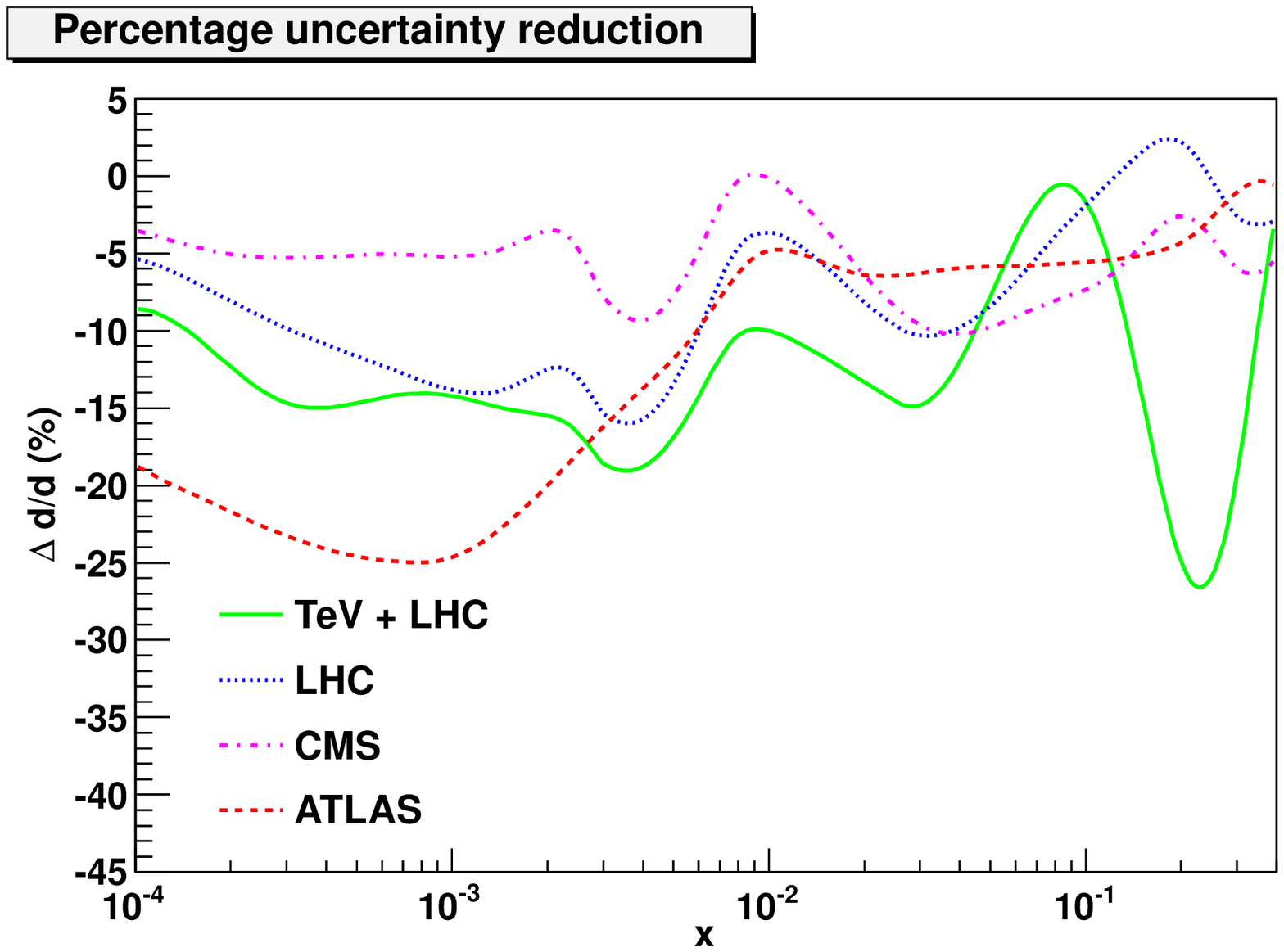,width=0.48\textwidth}
\epsfig{figure=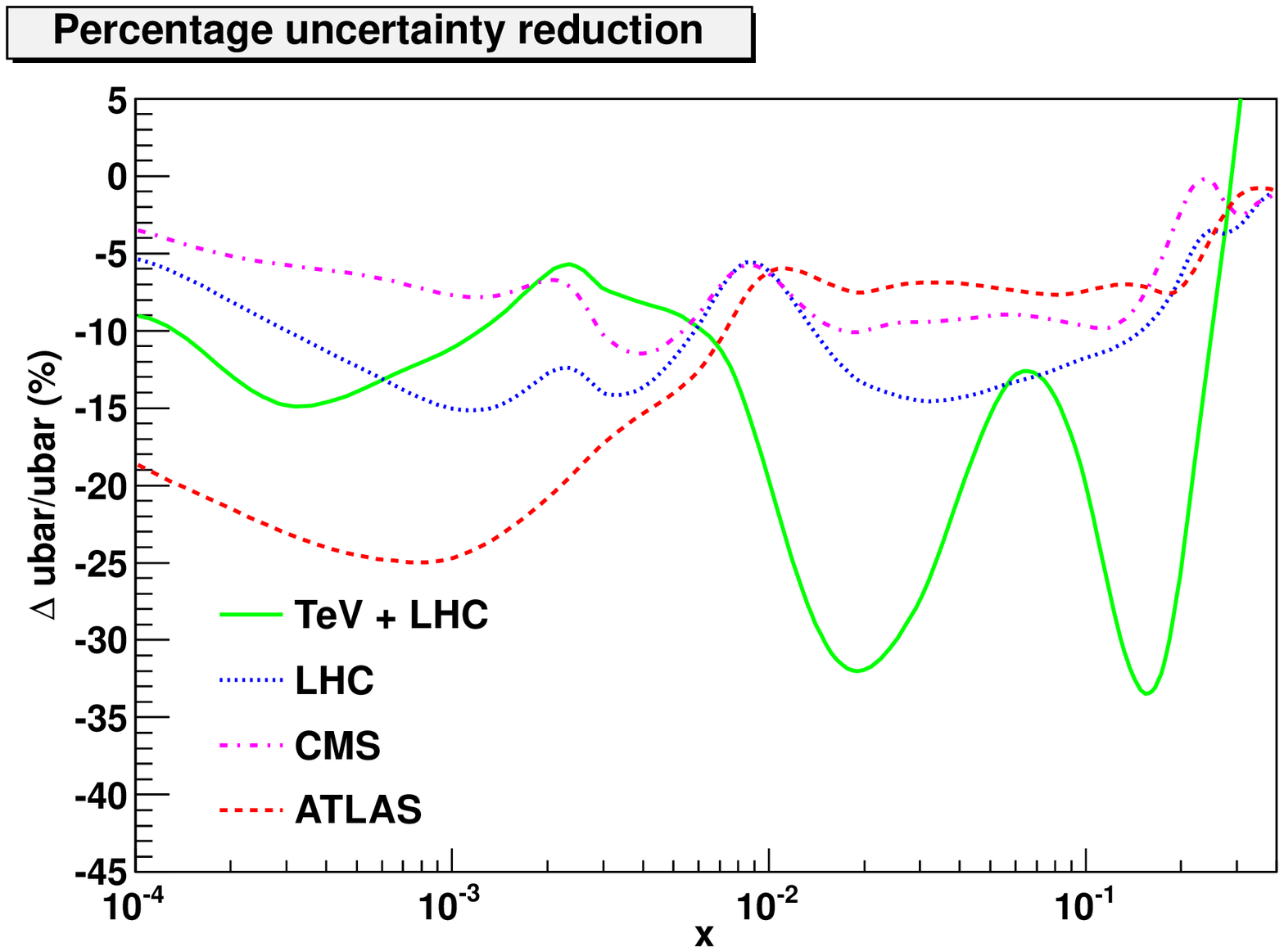,width=0.48\textwidth}
\caption{The percentage change in the uncertainty in the light quark and antiquark distributions at the scale $Q^{2}=M_{W}^{2}$ in the global NNPDF2.1 NNLO global fit, after adding ATLAS, CMS and D0 lepton charge asymmetry data via reweighting. The four curves show in each case the effect of ATLAS (red) and CMS (pink) only, together (blue), and then together with the D0 data (green).\label{fig:perc}}
\end{figure}

The description of the combined ATLAS, CMS and D0 charge asymmetry datasets obtained using the NNPDF2.1 NNLO global fit is better than the description obtained at NLO: $\chi_{\rm d.o.f.}^{\rm 2,NLO}$= 2.22 versus $\chi_{\rm d.o.f.}^{\rm 2,NNLO}$= 1.64. This is due to the NNLO corrections in the higher rapidity bins which bring the result closer to the experimental maesurements. After reweighting their overall description improves significantly, with a combined $\chi_{\rm d.o.f.}^{2}$= 0.97. This is due to a significant improvement in the fit to the D0 data (even though their description is still not optimal), as it is shown in Table~\ref{tab:chi2}. The fit to the ATLAS data deteriorates a little, showing that there is some tension. 
\begin{figure}
\epsfig{figure=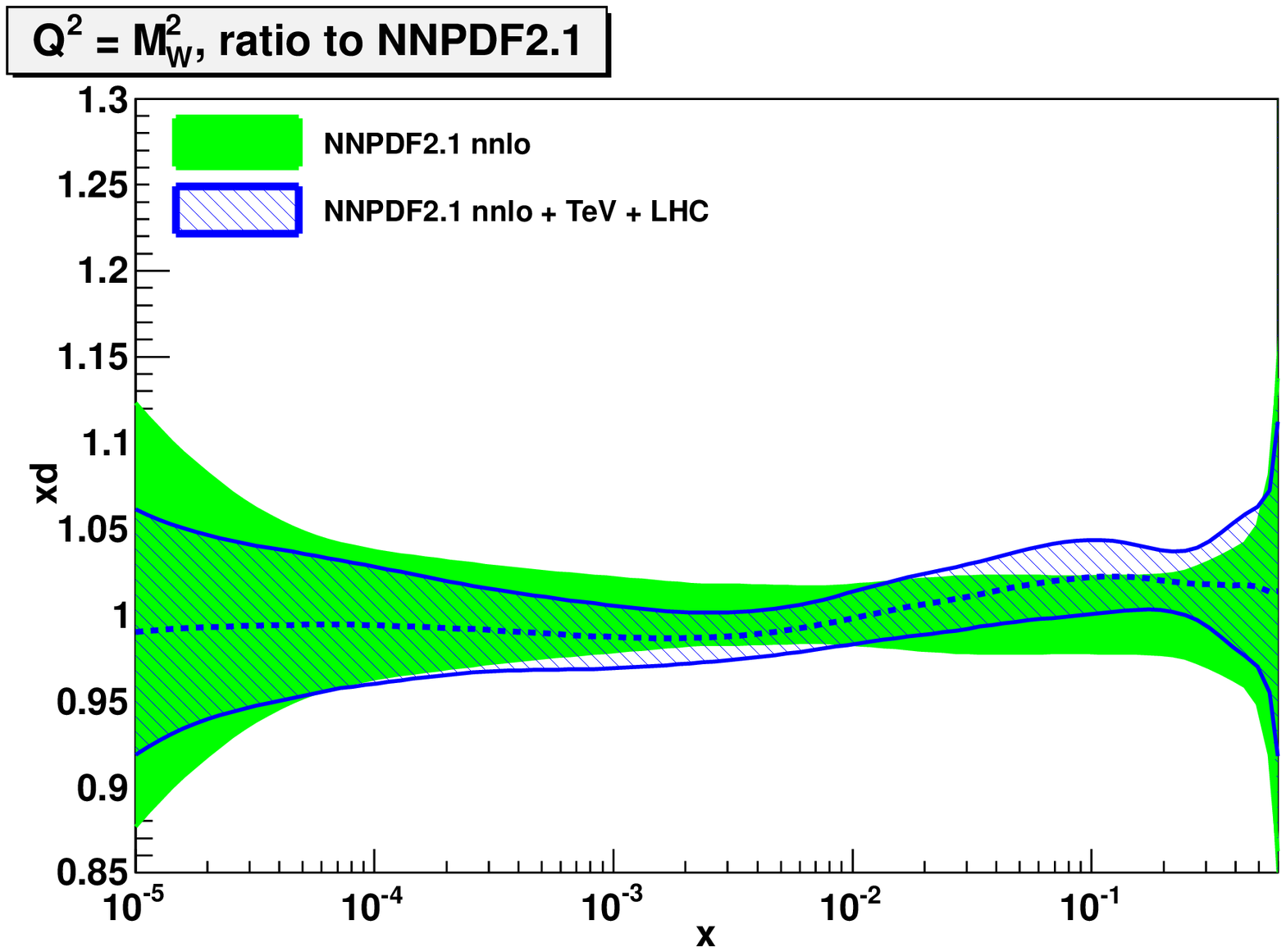,width=0.48\textwidth}
\epsfig{figure=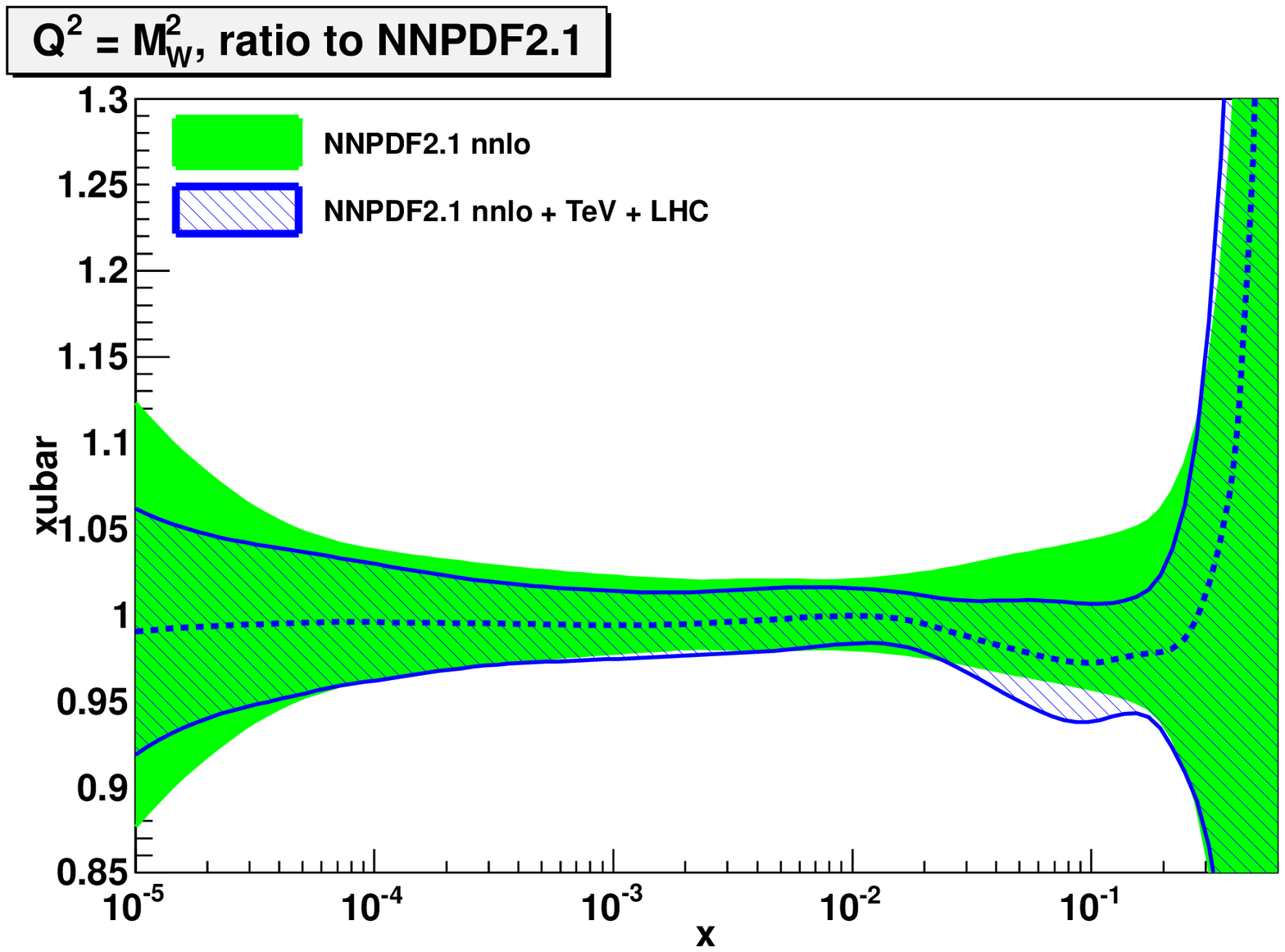,width=0.48\textwidth}
\caption{Same as Fig.~\ref{fig:pdfLHC} but after adding both the ATLAS+CMS and the D0 lepton charge asymmetry data.\label{fig:nnpdf22}}
\end{figure}
The number of effective replicas is now $N_{\rm eff}$ = 46 out of the initial $N_{\rm ref}$= 1000, indicating that the $W$ lepton asymmetry data indeed introduce very significant constraints on the PDFs. This is shown in Figs.~\ref{fig:nnpdf22} and \ref{fig:perc} where the light NNLO distributions are compared at the scale $Q^2=M_W^2$ to the ones obtained after reweighting the LHC and Tevatron data and the percentage reduction of uncertainty is displayed. The error reduction is concentrated in two separate regions of $x$, namely the $x\sim 10^{-3}$, which is mostly affected by the ATLAS data, and the $x\sim 10^{-2}-10^{-1}$ region, which is mostly affected by the CMS and D0 data. In each of these regions, the $W$ asymmetry data leads to a reduction of uncertainties on the light
flavour and antiflavour distribution, or around 25\% in the low $x$ region, and up to 30\% 
at higher $x$ when CMS and D0 are combined. In the latter region changes in the central values for these PDFs by up to 1$\sigma$ are also observed, mainly due to the D0 data.
\section{Conclusions}
The inclusion of the LHC 2010 and the Tevatron W lepton asymmetry data in a NNLO global parton analysis confirms the findings of the NLO analysis, showing that these data already provide some constraints on PDFs from LHC data. As the quantity and quality of LHC measurements potentially relevant for PDF determination increases at an impressive rate, this analysis is the first of a series of studies which which allow to assess the impact of the LHC data on PDFs as they come out.

\section*{Acknowledgments}
We thank the organizers of the Recontres de Blois for having organized a nice and stimulating workshop. MU is supported by the Bundesministerium f\"ur Bildung and Forschung (BmBF) of the Federal Republic of Germany (project code 05H09PAE). We would like to acknowledge the use of the computing resources provided by the Edinburgh Compute  and Data Facility (ECDF) (http://www.ecdf.ed.ac.uk/). The ECDF is partially supported by the eDIKT initiative (http://www.edikt.org.uk).

\end{document}